\documentclass[runningheads]{llncs}

\usepackage[colorlinks,linkcolor=red,anchorcolor=blue, citecolor=blue]{hyperref}
\usepackage{multirow}
\usepackage[T1]{fontenc}
\usepackage{graphicx}
\usepackage{cite}
\usepackage{subfigure}
\usepackage{tabularx}
\usepackage{float}
\usepackage{amsmath} 
\usepackage{amssymb}
\usepackage{booktabs}
\usepackage{todonotes}
\usepackage{color}
\usepackage{xcolor}
\usepackage{overpic}
\usepackage{subfigure}
\usepackage{bm}
\usepackage{bbding}
\usepackage{makecell}
\usepackage{array}
\usepackage{array}
\usepackage[T1]{fontenc}
\usepackage{graphicx}
\begin{document}
\title{ }
% \authorrunning{Caiwen Jiang et al.}
\title{A dual-task mutual learning framework for predicting post-thrombectomy cerebral hemorrhage}
\titlerunning{Prediction of post-thrombectomy cerebral hemorrhage.}

\author{Caiwen Jiang \inst{1,2} \and
Tianyu Wang \inst{4} \and
Xiaodan Xing \inst{2} \and
Mianxin Liu \inst{6} \and
Guang Yang \inst{2,7,8,9} \and
Zhongxiang Ding   \inst{4} \inst{(}\textsuperscript{\Envelope}\inst{)}\and
Dinggang Shen\inst{1,3,5}\inst{(}\textsuperscript{\Envelope}\inst{)}}
\institute{
School of Biomedical Engineering \& State Key Laboratory of Advanced Medical Materials and Devices, ShanghaiTech University, Shanghai, China\\ 
\email{\{jiangcw,dgshen\}@shanghaitech.edu.cn}\\
\and
Bioengineering Department and Imperial-X, Imperial College London, London, UK\\
\and
Shanghai United Imaging Intelligence Co., Ltd., Shanghai, China\\
\and 
Department of Radiology, Affiliated Hangzhou First People’s Hospital, Westlake University School of Medicine, Hangzhou, China \\
\and
Shanghai Clinical Research and Trial Center, Shanghai, China\\
\and
Shanghai Artificial Intelligence Laboratory, Shanghai, China\\
\and
National Heart and Lung Institute, Imperial College London, London, UK\\
\and
Cardiovascular Research Centre, Royal Brompton Hospital, London, UK\\
\and
School of Biomedical Engineering \& Imaging Sciences, King's College London, London, UK\\
}

\maketitle              % typeset the header of the contribution
%
% \vspace{-0.5cm}

\begin{abstract}
Ischemic stroke is a severe condition caused by the blockage of brain blood vessels, and can lead to the death of brain tissue due to oxygen deprivation. Thrombectomy has become a common treatment choice for ischemic stroke due to its immediate effectiveness. But, it carries the risk of postoperative cerebral hemorrhage. Clinically, multiple CT scans within 0-72 hours post-surgery are used to monitor for hemorrhage. However, this approach exposes radiation dose to patients, and may delay the detection of cerebral hemorrhage. To address this dilemma, we propose a novel prediction framework for measuring postoperative cerebral hemorrhage using only the patient's initial CT scan. Specifically, we introduce a dual-task mutual learning framework to takes the initial CT scan as input and simultaneously estimates both the follow-up CT scan and prognostic label to predict the occurrence of postoperative cerebral hemorrhage. Our proposed framework incorporates two attention mechanisms, i.e., self-attention and interactive attention. Specifically, the self-attention mechanism allows the model to focus more on high-density areas in the image, which are critical for diagnosis (i.e., potential hemorrhage areas). The interactive attention mechanism further models the dependencies between the interrelated generation and classification tasks, enabling both tasks to perform better than the case when conducted individually. Validated on clinical data, our method can generate follow-up CT scans better than state-of-the-art methods, and achieves an accuracy of $86.37\%$ in predicting follow-up prognostic labels. Thus, our work thus contributes to the timely screening of post-thrombectomy cerebral hemorrhage, and could significantly reform the clinical process of thrombectomy and other similar operations related to stroke.


\end{abstract} 

\section{Introduction}
\vspace{-1mm}
Ischemic stroke is a medical emergency caused by the blockage of blood vessels in the brain. Its timely treatment is crucial for reducing brain damage and other complications associated with stroke~\cite{shao2021new, jiang2024real}. Thrombectomy, favored for its quick effectiveness, has emerged as a common option for treating ischemic stroke~\cite{derex2017mechanical}. However, the procedure may damage blood vessels and require perfusion of contrast agents for vascular visualization, which could introduce a risk of postoperative cerebral hemorrhage.  In this context,  the timely screening of cerebral hemorrhage after thrombectomy is an essential clinical task.

In the clinic, cerebral hemorrhage is monitored by two to three CT scans conducted within 0-72 hours post-surgery~\cite{grkovski2022novel}. However, two to three CT scans cannot cover the entire period of cerebral hemorrhage (i.e., 0-72h post-surgery), often resulting in delayed detection, which can postpone the initiation of necessary treatment. Additionally, multiple CT scans within a short period also pose a significant radiation risk to patients. To address this dilemma, in this paper, we make the first attempt to predict the occurrence of hemorrhage within 0-72h post-surgery based only on the patient's initial CT scan.

There are already extensive studies on disease prediction~\cite{xie2021multi, hu2023vgg,alsekait2023toward, jiang2023s2dgan}. For example, Hu \textsl{et al} propose a framework that combines CNN and transformer for predicting the progression trends of mild cognitive impairment~\cite{hu2023vgg}. Alsekait \textsl{et al} integrate the support vector machine into deep learning models to predict the development of chronic kidney disease~\cite{alsekait2023toward}. However, these studies, which directly predict future prognostic labels from images, often lack intermediate evidence, rendering the prediction less convincing. Consequently, some studies attempt to achieve prediction by generating future images~\cite{han2022image, frid2018gan}. For instance, Han \textsl{et al} adopt the regularized generative adversarial networks to generate images of future time points for predicting the risk of osteoarthritis~\cite{han2022image}. Such approaches can provide more information, thereby making the outcomes more persuasive. In fact, estimating future prognostic labels and images does not conflict to each other, as there exists an inherent connection between the two tasks. Thus, we believe that performing both tasks simultaneously could potentially achieve better results than conducting them separately.

To this end, we design a dual-task interactive learning framework to simultaneously predict the follow-up CT scan and prognostic label from the patient's initial CT scan for achieving postoperative cerebral hemorrhage prediction. Through dual-task interactive learning, we can capture dependencies between the interrelated generation and classification tasks, allowing both tasks to perform better than the case when performed separately. Our proposed framework employs a combination of self-attention and interactive attention mechanisms. The self-attention mechanism enables the model to focus more on high-density areas that are critical for diagnosis. Meanwhile, the interactive attention mechanism models dependencies between the interrelated generation and classification tasks, significantly reducing computational complexity while enhancing the performance of each task. Extensive experiments on clinical data show that our method can generate higher-quality follow-up CT scans and achieve more accurate prognostic label prediction than state-of-the-art methods.

The main contributions of our work include \romannumeral1) the first attempt to achieve early prediction of postoperative cerebral hemorrhage by estimating follow-up CT scans and prognostic labels from initial scans, and \romannumeral2) the development of a novel dual-task interactive learning framework for this task. Extensive experiments also demonstrate the effectiveness of our method on collected datasets.

\begin{figure}[!t]
 \setlength{\abovecaptionskip}{0.1cm}
\setlength{\belowcaptionskip}{-0.4cm}
\centering
\begin{overpic}[width=1\linewidth]{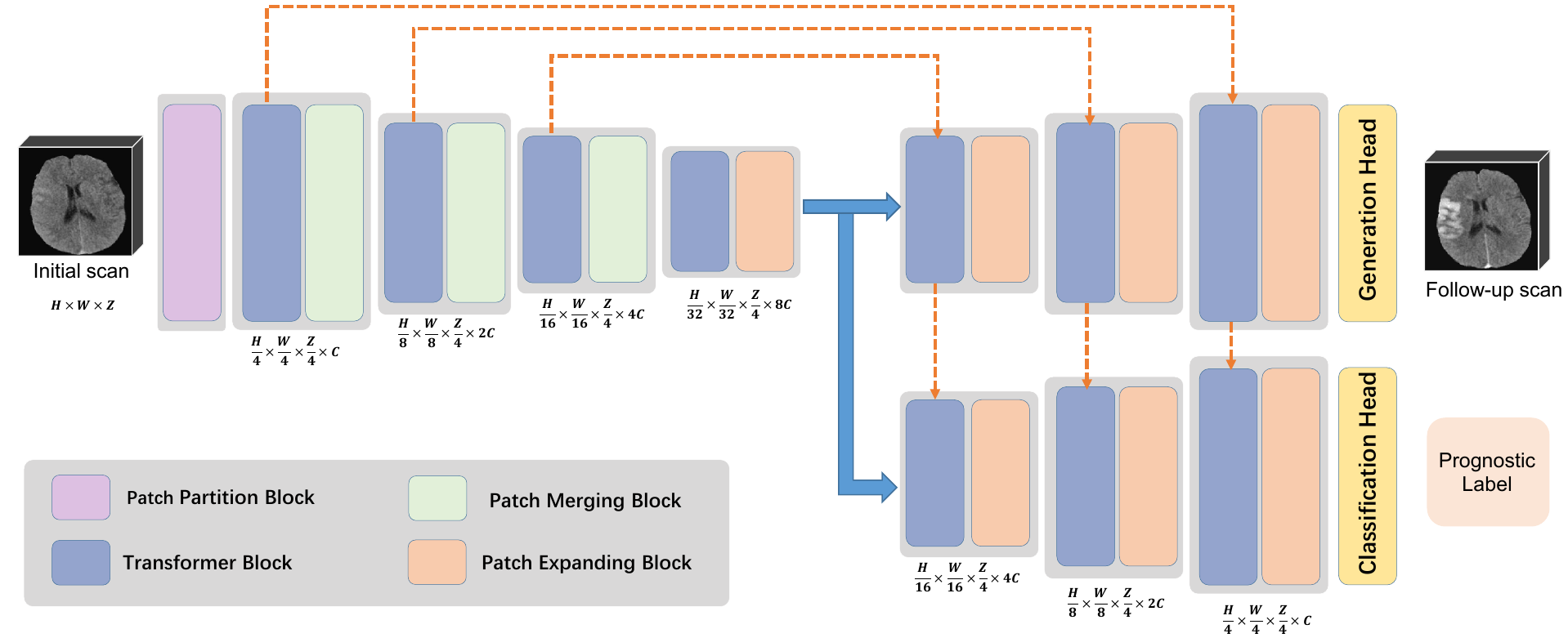}

    \end{overpic}
    \vspace{-4mm}
\centering
\caption{Overview of our proposed dual-task interactive learning framework.}
\label{fig1}
\end{figure}

\section{Method}
Our proposed dual-task interactive learning framework is shown in Fig.~\ref{fig1}. Given an initial CT scan, it is first processed by the patch partitioning block into a series of tokens that can be handled by subsequent transformer blocks. Then, these tokens alternately pass through transformer and patch merging blocks to extract features. The extracted features are subsequently input into two task-specific branches. Correspondingly, in each branch, the extracted features alternately pass through three transformer and patch expanding blocks to improve resolution. The final features are then fed into the corresponding task heads to predict the results. Throughout this whole process, in addition to the use of self-attention mechanism, we apply the interactive attention mechanism to perform attention interactions at the corresponding feature levels. In the following, we will introduce the details of our method.

\subsection{Spatial Alignment of Image Pairs}
\begin{figure}[!t]
 \setlength{\abovecaptionskip}{0.1cm}
\setlength{\belowcaptionskip}{-0.4cm}
\centering
\begin{overpic}[width=1\linewidth]{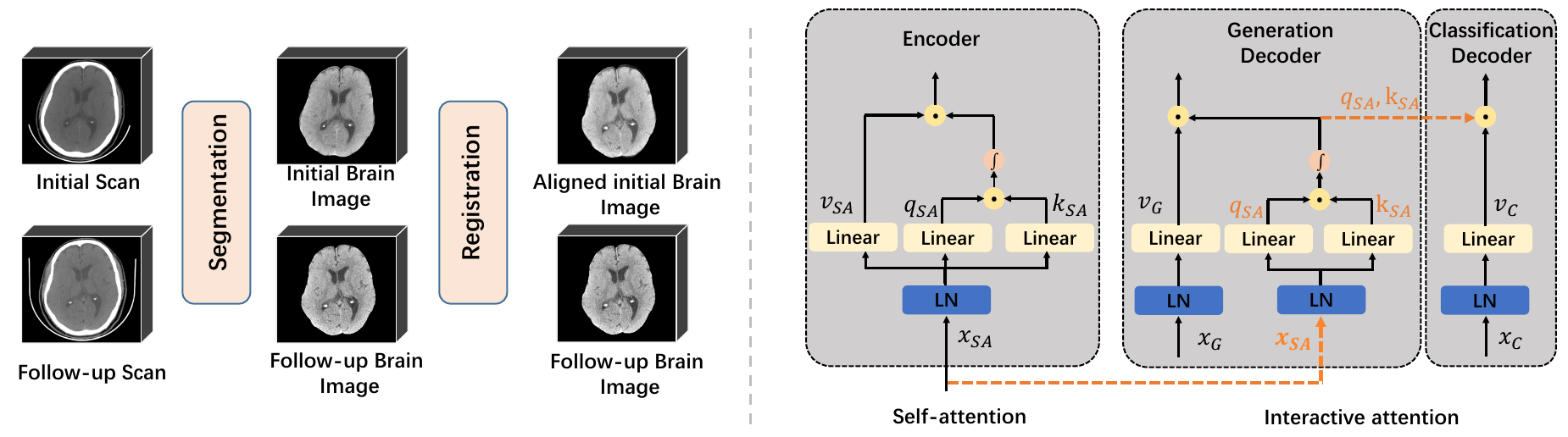}

    \end{overpic}
    \vspace{-4mm}
\centering
\caption{Left: Data preprocessing workflow for obtaining the spatially-aligned initial and follow-up brain images. Right: Details of the two attention mechanisms (i.e., self-attention, and interactive attention) involved in the proposed framework.}
\label{fig2}
\end{figure}

Due to patient posture and physiological movements, there is a significant spatial misalignment between initial and follow-up scans. Therefore, we need to perform data preprocessing to spatially align the initial and follow-up scans. As shown in the left part of Fig.~\ref{fig2}, for the input of two CT scans, to eliminate background interference, we first use the TotalSegmentator~\cite{wasserthal2023totalsegmentator}, an open-access tool based on nnU-Net and trained with more than one thousand samples, to segment the brain regions from both CT scans. Subsequently, we apply an affine registration method~\cite{avants2009advanced} to align the segmented brain regions. In this way, we can obtain spatially-aligned brain region image pairs for the latter model training.

\subsection{Model Architecture}
We propose a dual-task interactive learning framework consisting of five types of blocks, i.e.,  patch partitioning, transformer, patch merging, patch expanding, and task head. Among them, the patch partitioning block splits the input into multiple non-overlapping patches, with the features of each patch being the concatenation of the raw voxel values. 

For the transformer blocks, we use the same window operation as Swin-transformer~\cite{liu2021swin}, i.e., computing attention in the partitioned windows, instead of the whole images or feature maps. Specifically, each transformer block contains a regular window-based multi-head self-attention (W-MSA) module and a shifted window-based MSA (SW-MSA) module, followed by a 2-layer multilayer perceptron (MLP). Layer normalization (LN) is applied before each MSA module and MLP layer, and residual connections are applied after each module.

Patch merging and patch expanding blocks can be regarded as two opposite operations. The patch merging block merges adjacent tokens along the height and width dimensions in a non-overlapping manner to generate new tokens. In implementation, our merging scope is $2 \times 2$; therefore, after passing the patch merging layer, the height ($H$) and width ($W$) dimensions of the features are halved, and the $C$ dimension is quadrupled. Then, a linear mapping is applied to halve the channel dimension of the concatenated tokens. Correspondingly, the patch expanding block first doubles the channel dimension of the input features through linear mapping, and then reshapes the features to double the height and width dimensions while reducing the channel dimension.

The task-specific heads are used to predict the corresponding task results from the features. The generation head and classification head are each composed of a single linear layer and a single softmax layer, respectively. We employ a weighted sum~\cite{chen2018gradnorm} to dynamically adjust the training weights of each task-specific loss according to their gradients. The task-specific loss is calculated between the ground truth and the final predictions for each task. In particular, we use both $L1$ loss and adversarial loss for generation task, and a cross-entropy loss for classification task.

% \vspace{-2mm}
\subsection{Self-Attention and Interactive Attention}
Due to the higher density of blood compared to normal brain tissue, cerebral hemorrhage regions typically appear as high-density areas in CT images. To better extract features from CT images that may contain high-density areas, our proposed framework employs two attention mechanisms, including self-attention and interactive attention, as shown in the right part of Fig.~\ref{fig2}.

\textsl{Self-Attention:} The self-attention is executed during feature extraction. By computing self-attention in the encoder, we enable the model to focus more on high-density areas in the image, which are critical for diagnosis. Specifically, the input $x_{_{SA}}$ first passes through three linear layers to obtain query $q_{_{SA}}$ and key $k_{_{SA}}$. Then, the attention value $A_{_{SA}}$ is calculated as:
\begin{equation}
A_{_{SA}} = softmax(\frac{q_{_{SA}}k_{_{SA}}}{\sqrt{C_{_{SA}}}}+B),
\label{equation 2}
\end{equation}
where $C_{_{SA}}$ is the number of channels and $B$ is the position bias. Finally, the output of a particular self-attention head is $A_{_{SA}}\cdot v_{_{SA}}$.  In this way, we can apply adaptive weights to different areas of the image or feature map, allowing the model to focus more on high-density areas.

\textsl{Interactive Attention:} It is known existing strong correlation between follow-up CT scans and prognostic labels. For example, follow-up CT scans can be used to diagnose prognostic labels, and prognostic labels can roughly describe follow-up CT scans. Therefore, the task of generating follow-up CT scans and the task of predicting prognostic labels should also be interrelated. To capture task dependencies beyond shared encoder parameters, we design an interactive attention mechanism in decoders to further capture the relationship between these two tasks, reducing computational overhead while enhancing the performance of both tasks.

In our implementation, we set the generation task as the reference task. For a specific interactive attention calculation in the reference task decoder (i.e., the generation decoder), let $x_{_{G}}$ denote the previous block output, and $x_{_{SA}}$ denote the output of the corresponding transformer block in the encoder. As shown in the right part of Fig.~\ref{fig2}, the generation decoder takes both $x_{_{G}}$ and $x_{_{SA}}$ as input. The standard method of computing self-attention is to obtain key, query, and value vectors only from its own previous output $x_{_{G}}$. In contrast, in the interactive attention calculation, we compute the query $q_{_{SA}}$ and key $k_{_{SA}}$ from $x_{_{SA}}$ (from the encoder). Meanwhile, the value $v_{_{G}}$ is still computed using the previous block output $x_{_{G}}$ since the final output should be related to the generation task. 

For the classification decoder, we adopt the same scheme as above, but we only calculate $A_{_{SA}}$ in the decoder of reference task (i.e., generation task), and then feed it to the classification decoder directly. Note that we can also take the classification task as the reference task. However, we find that, taking the generation task as the reference task can lead to better outcomes through experiments. We apply the same procedure to all transformer blocks in the generation and classification decoders.

\section{Experiments}
\vspace{-2mm}
\subsection{Dataset and Implementation}
We collect 200 samples for our dataset, and each sample contains the initial CT scan, the final follow-up CT scan, and the follow-up prognostic label (i.e.,  hemorrhagic transformation and non-hemorrhagic transformation). The follow-up prognostic label is determined by doctors based on the final follow-up CT scan. Of these 200 scans, 160 are used for training and 40 for testing. During the evaluation, we conduct five-fold cross-validation to exclude randomness.

In our implementation, experiments are conducted on the PyTorch platform using two NVIDIA Tesla A100 GPUs and an Adam optimizer with an initial learning rate of $0.001$. All images are resampled to a voxel spacing of $1\times1\times1~\text{mm}^{3}$ with the size of $256 \times 256 \times 128$, and their intensity is normalized within [$0,1$] by min-max normalization. To augment the training samples and reduce the usage of GPU memory, the original image is randomly cropped to the size of $96 \times 96 \times 96$ as input. To quantify our results, we use Peak Signal to Noise Ratio (PSNR) and Structural Similarity Index Measure (SSIM)~\cite{jiang2023semi} to evaluate the generation task, and ACCuracy (ACC) and Area Under the Curve (AUC) to evaluate the classification task.

\begin{table}[!t]
\centering
\caption{Quantitative results of ablation analysis, in terms of PSNR, SSIM, ACC, and AUC.}
\vspace{-3mm}
\begin{tabular}{l|cc|cc} 
\toprule
\multirow{2}{*}{Method} & \multicolumn{2}{c|}{\textbf{Generation}} & \multicolumn{2}{c}{\textbf{Classification}}  \\ 
\cline{2-5}
                          & PSNR $[dB]\uparrow$ & SSIM $[\%]\uparrow$       & ACC $[\%]\uparrow$ & AUC $[\%]\uparrow$                        \\ 
\hline
S-CNN                     & 22.28(2.31)   & 86.45(2.37)      & 79.28(2.93)   & 82.43(3.12)                          \\
D-CNN                    & 25.53(1.26)   & 89.37(2.13)      & 82.36(2.85)   & 84.92(3.22)                          \\ 
\hline
S-Transformer              & 25.47(1.23)   & 90.42(1.18)      & 83.64(2.13)   & 86.54(2.64)                          \\
D-Transformer              & 26.92(1.18)   & 92.17(1.13)      & 85.23(1.95)   & 89.75(2.05)                         \\
Ours                  & $\bm{28.57 (1.02)}$   & $\bm{92.48 (1.12)}$   &  $\bm{86.37(1.84)}$   & $\bm{92.32(2.14)}$                         \\
\bottomrule
\end{tabular}
\label{table1}
\end{table}

\subsection{Ablation Analysis}
To evaluate the effectiveness of each network component in our dual-task mutual learning framework, we designed another four variants, including: 1) S-CNN, consisting of a CNN encoder and a CNN decoder; 2) D-CNN, consisting of a CNN encoder and two CNN decoders; 3) S-Transformer, consisting of a transformer encoder and a transformer decoder; 4) D-Transformer, consisting of a transformer encoder and two transformer decoders. D-Transformer has the same architecture as our method, but without adopting the interactive attention mechanism in decoders. In addition, for S-CNN and S-Transformer, we need to use two separate models to perform the generation and classification tasks, respectively.

The quantitative results are provided in Table~\ref{table1}, from which, we can find the following observations. (1) D-CNN/D-Transformer achieves better performance than S-CNN/S-Transformer. This proves that a dual-task learning framework is more appropriate than a single-task framework for our interrelated generation and classification tasks. (2) The transformer-based S-Transformer and D-Transformer, respectively, achieve better results than the CNN-based S-CNN and D-CNN. This may be because the transformers can capture global information by focusing on high-density areas crucial for diagnosis, thereby benefiting both tasks. (3) Our method achieves better results on both generation and classification tasks than D-Transformer and other variants. This demonstrates that the interactive attention mechanism can strengthen the connection between generation and classification tasks, thus resulting in better performance. These three comparisons conjointly verify the effective design of our proposed framework, where the \textsl{dual-task learning strategy}, \textsl{transformer-based architecture}, and \textsl{interactive attention mechanism} all can benefit our tasks.

\begin{table}[!t]
\centering
\caption{Quantitative comparison of our method with several state-of-the-art generation and classification methods, in terms of PSNR, SSIM, ACC, and AUC, where $^{\ast}$ denotes CNN-based method and $^{\dagger}$ denotes Transformer-based method.}
\vspace{-3mm}
\begin{tabular}{l|cc|l|cc} 
\toprule
% \multirow{2}{*}{Method} & \multicolumn{2}{c|}{\textbf{Generation}} &\multirow{2}{*}{Method}& \multicolumn{2}{c}{\textbf{Classification}} \\ 
% \cline{2-3} 
%                           & PSNR$\uparrow$ & SSIM$\uparrow$  &                       & PSNR$\uparrow$ & SSIM$\uparrow$                               \\ 

\multicolumn{3}{c|}{\textbf{Generation}}& \multicolumn{3}{c}{\textbf{Classification}}\\
\hline
Method  & PSNR $[dB]\uparrow$ & SSIM $[\%]\uparrow$     & Method   & ACC $[\%]\uparrow$ & AUC $[\%]\uparrow$                                \\      
\specialrule{0em}{1pt}{0pt}                          
\hline
cGAN$^{\ast}$~\cite{isola2017image}     & 23.32(1.75)   & 85.44(1.86) &   VGG$^{\ast}$~\cite{mahjoubi2023deep}   & 79.46(3.75)   & 82.12(2.97)                               \\
SAGAN$^{\ast}$~\cite{lan2021three}      & 24.84(1.43)   &88.34(1.54) &   ResNet$^{\ast}$~\cite{zhou2022transfer}   & 81.46(4.12)   & 85.76(3.46)                                \\ 
\hline
TransUNet$^{\dagger}$~\cite{chen2021transunet}     & 26.12(1.22)  & 89.45(1.26) &  Trans-RNN$^{\dagger}$~\cite{ayoub2023end}   & 83.02(2.17)   & 87.96(2.56)                                \\ 

ResViT$^{\dagger}$~\cite{dalmaz2022resvit}  &  27.34(1.23)   & 89.73(1.14) & Res-Trans$^{\dagger}$~\cite{wang2022vision} & $84.66(2.43)$   &  $89.42(3.12)$                              \\   

Ours$^{\dagger}$           & $\bm{28.57 (1.02)}$   & $\bm{92.48 (1.12)}$ &   Ours$^{\dagger}$   &  $\bm{86.37(1.84)}$   & $\bm{92.32(2.14)}$                                  \\
\bottomrule
\end{tabular}
\label{table2}
\end{table}

\begin{figure}[!t]
 \setlength{\abovecaptionskip}{0.1cm}
\setlength{\belowcaptionskip}{-0.4cm}
\centering
\begin{overpic}[width=1\linewidth]{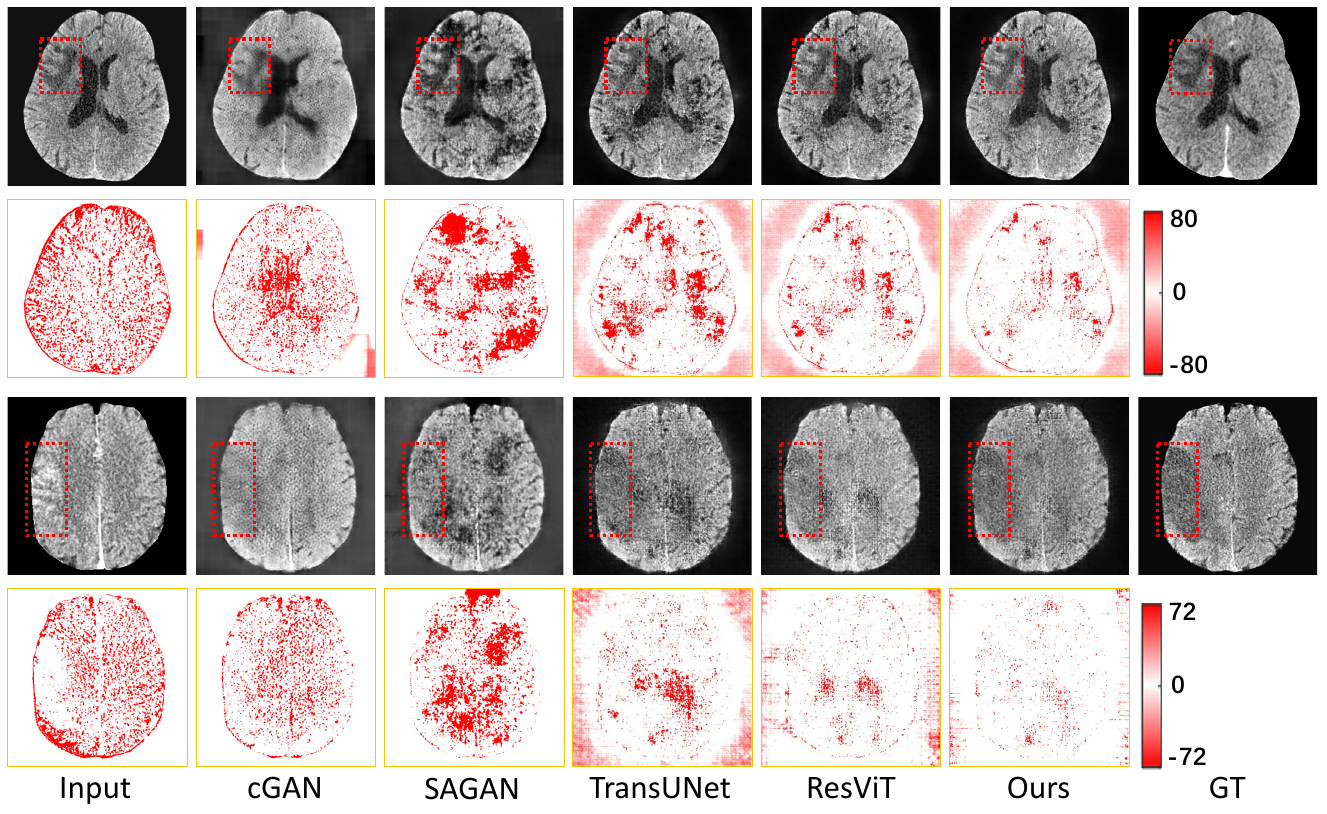}
   
    \end{overpic}
    \vspace{-5mm}
\centering

\caption{Visual comparison of follow-up scans produced by five different methods. From left to right are the input (initial scan), results by five other comparison methods (2nd-5th columns) and our method (6th column), and the ground truth (GT, i.e., the follow-up scan). The corresponding difference maps between the generated results and GT are shown in the 2nd and 4th rows, where darker color indicates larger differences. Red dotted boxes show the areas for detailed comparison.}
% \vspace{-0.3cm}
\label{fig3}
\end{figure}

\subsection{Comparison with State-of-the-art Methods}
Furthermore, we compare our method with several state-of-the-art generation and classification methods. The generation methods include cGAN~\cite{isola2017image}, SAGAN~\cite{lan2021three}, TransUNet~\cite{chen2021transunet}, and ResViT~\cite{dalmaz2022resvit}. The classification methods include VGG~\cite{mahjoubi2023deep}, ResNet~\cite{zhou2022transfer}, Transformer-RNN (Trans-RNN)~\cite{ayoub2023end}, and ResNet-Transformer (Res-Trans)~\cite{wang2022vision}. The quantitative results and visualizations of the generated outcomes are provided in  Table~\ref{table2} and Fig.~\ref{fig3}, respectively.

\vspace{1mm}
\textsl{Quantitative Comparison:} 
Quantitative results are provided in Table~\ref{table2}. It can be observed that, overall, transformer-based methods outperform CNN-based methods on both generation and classification tasks. This may be attributed to the transformer structure's superior ability to extract and focus on high-density areas crucial for diagnosis. This validates our selection of the transformer-based architecture. Further, among all the transformer-based methods, our method achieves the best performance. This demonstrates that employing a dual-task framework to simultaneously perform interrelated generation and classification tasks yields better performance than performing any of those tasks individually.

\vspace{1mm}

\textsl{Qualitative Comparison:}
We provide a visual comparison of follow-up scans generated by five different methods in Fig.~\ref{fig3}. First, compared to other methods, our method can generate the overall optimal images, characterized by the least noise, fewest artifacts but clearest structure.  Second, in terms of detail, our method can also most accurately generate the high-density areas (i.e., areas marked by red boxes) that are crucial for predicting cerebral hemorrhage. Finally, the lightest color in the difference map demonstrates our method can generate lung images with the smallest difference from the ground truth. Such key observations demonstrate that our method is superior to those state-of-the-art methods in generation task.

% \vspace{-2mm}
\section{Conclusion}
In this paper, to preemptively determine the occurrence of cerebral hemorrhage post-thrombectomy, we have presented a novel prediction method based solely on the patient's initial CT scan, i.e., simultaneously predicting the follow-up CT and prognostic label from the initial scan. To achieve this goal, we design a dual-task mutual learning framework by proposing three novel strategies including 1) dual-task learning strategy, 2) transformer-based architecture, and 3) interactive attention mechanism. Among them, the transformer-based architecture enables the model to focus more on the areas important for diagnosing cerebral hemorrhage. Dual-task learning strategy and interactive attention mechanism capture the dependencies between the interrelated generation and classification tasks to improve performance while effectively reducing computational complexity. Validated on the collected clinical dataset demonstrates that our method is designed effectively and can achieve superior performance quantitatively and qualitatively over the state-of-the-art methods.

\subsubsection{\ackname} This work was supported in part by National Natural Science Foundation of China (grant numbers U23A20295, 62131015, 62250710165), the STI 2030-Major Projects (No. 2022ZD0209000), Shanghai Municipal Central Guided Local Science and Technology Development Fund (grant number YDZX20233100001001), the China Ministry of Science and Technology (STI2030-Major Projects-2022ZD0213100), The Key R\&D Program of Guangdong Province, China (grant numbers 2023B0303040001, 2021B0101420006), the ERC IMI (10100\\5122), the H2020 (952172), the MRC (MC/PC/21013), the Royal Society (IEC\textbackslash NS\\FC\textbackslash211235), the NVIDIA Academic Hardware Grant Program, the SABER project supported by Boehringer Ingelheim Ltd, NIHR Imperial Biomedical Research Centre (RDA01), Wellcome Leap Dynamic Resilience, UKRI guarantee funding for Horizon Europe MSCA Postdoctoral Fellowships (EP/Z002206/1), and the UKRI Future Leaders Fellowship (MR/V023799/1).

\subsubsection{Declaration of Competing Interest.}
The authors declare that they have no known competing financial interests or personal relationships that could have appeared to influence the work reported in this paper.

{\small 
\bibliographystyle{unsrt} 
\bibliography{myref.bib} } 

\end{document}